\documentclass{article} 

\usepackage{graphicx} 
 
\begin{document} 
 
\begin{center} 
{\bf \Large Off-lattice simulation of the solid phase DNA amplification}\\[5mm]
{\em  {\large M. J. Krawczyk$^*$ and K. Ku{\l}akowski$^\dagger$}\\[3mm]   
Department of Applied Computer Science,  Faculty of Physics and Applied Computer Science
,  AGH University of Science and Technology\\ 
al. Mickiewicza 30, PL-30059 Krak\'ow, Poland.\\[3mm] 
 
$^*$E-mail:
$^*$krawczyk@novell.ftj.agh.edu.pl,
$^\dagger$kulakowski@novell.ftj.agh.edu.pl\\[3mm]    \today
}
\end{center} 
 
\begin{abstract} 
Recent simulations of the solid phase DNA amplification (SPA) by
\mbox{J.-F. Mercier} et al (Biophys. J. 85 (2003) 2075) are generalized to
include two kinds of primers and the off-lattice character of the primer
distribution on the surface. The sigmoidal character of the primer occupation
by DNA, observed experimentally, is reproduced in the simulation. We discuss
an influence of two parameters on the efficience of the amplification process:
the initial density $p_0$ of the occupied primers from the interfacial
amplification and the ratio $r$ of the molecule length to the average
distance between primers. The number of cycles till the saturation decreases
with $p_0$ roughly as $p_0^{-0.26}$. For $r=1.5$, the number 
of occupied primers is reduced by a factor two, when compared to the case 
of longer molecules. Below $r=1.4$, the effectivity of SPA is reduced 
by a factor $100$.
\end{abstract}

\section{Introduction}  

The standard polymerase chain reaction (PCR) allows to produce multiple
copies of DNA {\it in vitro}. It is known as a revolutionary technique for
molecular biology \cite{wolfe}. The reaction takes place in the whole volume
of a vial. If different molecules of DNA are to be amplified, the reaction
products must be separated by other methods \cite{msma}. This separation is successfully
omitted in a recently introduced technique of the solid phase DNA amplification
\cite{bing,adessi}, which can be classified as an application of the
emerging biochip technology \cite{sch,che}.

The technique, termed as SPA, contains three stages. At first, primers of two
kinds are attached by their 5' ends to a solid surface. They are to be
complementary to two ends of the investigated DNA strands, then their
proportions should be 1:1. Next, small amount of DNA is attached to some of
the primers in one thermal cycle. This stage is termed as the interfacial
amplification (IA). At third stage (surface amplification, SA), the solution of DNA is
washed out, and it is only the attached strands which are multiplicated in
subsequent cycles. They attach their free ends to neighbouring primers. Next,
they are copied by standard PCR. One thermal cycle can be separated in three distinct steps - annealing, extension and denaturation, which are repeated in an iterative way. In
this way, the concentration of occupied primers (i.e. primers with attached
DNA) increases. In this final stage of SPA the area of occupied primers widens
by a slow motion of its borders: the border velocity is not larger than the
strand length per cycle. A detailed and transparent description of the process
can be found in \cite{msma}, together with first Monte Carlo simulations.

The aim of this work is to investigate numerically the range of experimental 
parameters, where the surface amplification is effective. Our contribution can 
be treated as a
computational supplement to Ref. \cite{msma}. However, there are two main differences
between our approach and Ref. \cite{msma}. First, our simulation is
performed in the off-lattice scheme, i.e. the positions of the primers on the
surface do not form a square lattice, but they are randomly distributed.
Second, we take into account the fact that there are primers of two kinds. 
Both modifications are 
introduced to make 
the simulation more realistic. In particular, we intend to capture the limitation 
of SPA which arise when the length of the strands used is of the order of the 
mean distance between the primers. The distance distribution between primers is 
deformed by the assumption of the square lattice, and this deformation is 
particularly strong for short distances. Here this assumption is omitted. 
On the other hand, once a molecule of DNA is attached to a
given kind of primer by one end, its free end can only be attached to the
primer of the other kind. If the distance between primers is of the order of
the molecule length, this condition additionally limits the surface
amplification process. Again, here this limitation is captured properly.

We also investigate an influence of other parameters: of the efficiency of
the interfacial amplification and
of the mutual proportion of two kinds of primers. The
results can be useful for designing the range of parameters, where SPA is
optimal. If the concentration of occupied primers after the interfacial
amplification is low, the number of SPA cycles must be large, what raises 
costs. Note that this concentration is connected directly with the
density of the investigated DNA in the sample, and we are obviously
interested in the possibility of detecting low amounts of DNA. 

In the next section we discuss the values of the input parameters for the 
calculation.  One of them, the density $\rho _0$ of occupied primers at the beginning
of the simulated process, is the output of IA and therefore it can be treated 
as a measure of its efficiency. In this section we describe also the simulation 
method. In subsequent section our numerical results are reported. The text is 
closed by discussion. 

\section{Calculations}

In our computational model we refer to the data on SPA given in Ref. \cite{adessi}.
There, an optimal value of the density of the primers on a surface has been
found to be $15 fmol/mm^2$. After two cycles
of IA and SA, the density $\rho _0$ of occupied primers is $0.0011 fmol/mm^2$, i.e. a 
molecule of DNA is attached to one per 15000 primers. This amount comes mainly from IA 
\cite{adessi}. 
Subsequent 28 cycles of SA 
lead to an increase of $\rho$ to $0.0029 fmol/mm^2$. If the process of SPA is performed 
without washing the sample, IA and SA occur simultaneously. In this case, 28 cycles 
lead to $\rho =0.019 fmol/mm^2$. 

These data allow to estimate the efficiency of both IA and SA at one cycle of SPA, 
for the case when $\rho $ is small enough. This condition, true at few initial cycles, 
allows to neglect the slowing down of the amplification because of the overcrowding
by occupied primers. Then, the density $\rho$ increases in subsequent cycles of SA by a 
constant factor $\alpha $, and $\rho _n=\alpha ^n\rho _0$. From the data given 
above we deduce that $\alpha ^{28}=29/11$, 
i.e. $\alpha =1.035$. For such a small rate of amplification, the increase of $\rho$ 
in first two cycles of IA+SA is due mainly to IA. As we expect that this increase is 
linear with the number of cycles, we get its value equal to 
$x=\rho _0/(1+\alpha )=0.00055 fmol/mm^2$ per cycle.

For small $\rho$, at each cycle of IA+SA the density of occupied primers should increase 
according to the rule: 

\begin{equation}
\rho _{n+1}=x+\rho _n*\alpha
\end{equation}

Applying this rule again 28 times with $\rho _0=0.0011 fmol/mm^2$, we get 
$\rho _{28} =0.029 fmol/mm^2$, which is about 50
percent larger, than the experimental value. We deduce that this value cannot be treated
as small, and the effect of a local overcrowding should be taken into account. 
This means, that Eq.(1) cannot be applied: a simulation is necessary.

The area of the surface used for SPA is about 1 $mm^2$, and the number of primers $pN$
there is equivalent to 15 $fmol$, i.e. about $9*10^9$ primers. This amount is too large 
for our computational resources. We work with $K=15000$ primers, which is equivalent to 
the area $S$ about 1 $\mu m^2$ covered with the density 15 $fmol/mm^2$. To improve 
statistics, we average the results over $k$ runs.
The positions of the primers are selected randomly with a constant probability 
distribution on a squared area with periodic boundary conditions. Then, DNA strands
is attached to some primers. The amount of these primers is a measure of the 
effectiveness of IA. Simultaneously, it is proportional to the starting value of 
$p$ for the simulation of SA. The amount of occupied primers is 

\begin{equation}
\rho S=pK
\end{equation}

The density value $\rho =0.0011 fmol/mm^2$ is equivalent to $p=1/15000$, 
i.e. one occupied primer as a starting point for the simulation.

Two kinds of primers fit to two ends of the investigated molecules of DNA. Let us denote 
the primers by $\pm 1$. For each occupied primer $i$, the algorithm of the simulation of 
SA selects a primer $j$ which is {\it i)} its closest neighbour {\it ii)} of opposite sign
and {\it iii)} it is free. The 
additional condition is that the distance between $i$ and $j$ cannot excess the 
molecule length. During each cycle of SPA, the molecule attached to $i$ with one end is 
attached with another end to $j$ with probability depending on the distance according to 
the Gaussian function, multiplied by a phenomenological factor $\gamma $. The choice 
of the Gaussian function reflects the random-walk character of the free end of the 
molecule. The factor $\gamma $ is set as to reproduce the experimental value of IA, 
expressed by $\alpha =1.035$. This value means that during one cycle of SPA, only a few 
percent of DNA strands are amplified .
Our algorithm is approximate, because we neglect the possibility that the free molecule
end is attached to another primer, which is not the closest one. This 
simplification is introduced to speed up the calculations. According to the Gaussian 
distribution, the attaching probability strongly decreases with the distance. That is why
this approximation seems to be acceptable.

\section{Results}

In Fig.1 we show the ratio $p$ of the number of occupied primers to their
total number against time, the latter expressed in the number $N$ of thermal
cycles. The starting value of $p$ is taken as $p_0=1/15000$, and it is due to the
amount of primers occupied in the interfacial amplification. The increase of
$p$ shown in the picture is a consequence of the surface amplification. The
curve is obtained by averaging over $14$ simulations. The error bars are due
to the statistics. The sigmoidal curve  allows to evaluate the overall dynamics of the 
process of SA.

In Fig. 2, the number $N$ of cycles needed to obtain the saturation $p=1$ is shown as
dependent on the number of primers $p_0$ occupied by means of the interfacial
amplification. As we see, $N$ decreases as $p_0^{-\phi}$, at least for $p_0$
far from the target value $p=1$. We get $\phi $=0.264. Note that the data in Fig. 2 
are not equivalent to 
those in Fig. 1. To obtain each point of the plot, a separate simulation is performed.
During the process of SA, spatial correlations between occupied primers develop
from an initially random configuration.  

In Fig. 3 we show the number $N$ of cycles needed for the saturation, against
the ratio $r =L/d$, where $L$ is the molecule length and $d=(S/K)^{1/2}$ 
is the mean distance 
between primers. The plot shows a plateau for $r > 1.8$; below 
this value, $N$ abruptly increases. In Fig. 4 the limit value of $p$ is shown,
which stabilizes after many steps of SA, as dependent on the ratio $r$.
As we see, below $r =1.4$, the process of SA does not work.

In Fig. 5 we show how SPA is sensitive to a variation of the ratio $c$ of the
numbers of two kinds of the primers. Obviously, an optimum of the technique is
for $c=0.5$, when we have in the average the same numbers of the primers
appropriate for both ends of the investigated molecules of DNA. However, the
method works also for other values of $c$. We see that for $r=2$, the
saturated state $p=1$ can be reached for the range of $c$ as wide as
$(0.2,0.8)$. This means, that for shorter molecules or for a smaller density of
primers, the proportion of two kinds of primers should be controlled more
precisely.

\section{Discussion}

The exponential character of the results presented in Fig. 2 allow to evaluate the
time of measurement by SA as dependent on $p_0$. The value of $p_0$ is directly proportional
to the efficiency of the interface amplification, which depends both on the time of IA and 
on the concentration
of DNA in investigated samples. Further, the results shown in Fig. 3 prove, that the 
efficiency of the 
surface amplification depends on the molecule length only if the latter is of order 
of the distance $d$ between primers. The same conclusion can be drawn from Fig.4. 
We note that in this case, $d$ can be evaluated as $\rho ^{1/2}$. These results provide a direct
information on the range of parameters where the technique of the surface amplification can be used.
Also, the obtained limitations of the range of the parameter $c$ can be of interest for
applications. Actually, this result show that the technique works quite well even if the 
proportion of two kinds of primers is between 1:4 and 4:1.

A drawback of our simulation is that interactions between the molecules of DNA is not taken
into account. This interaction is known to prevent the saturation; the obtained values
of $p$ stabilize near 1/300 instead of 1.0 \cite{adessi}. This is a severe limitation
of the efficiency of the whole technique. However, we expect that with an appropriate rescaling 
of the value of the parameter of $p$ at saturation, our results remain qualitatively correct.

\begin{figure}[ht]
\begin{center}
\includegraphics[angle=-90,width=.9\textwidth]{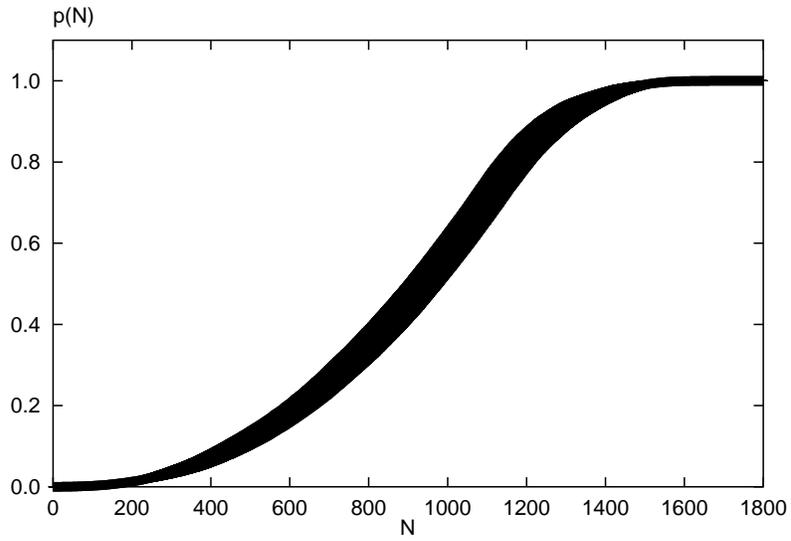}
\caption{Fraction of occupied primers $p$ against the cycle number $N$.}
\label{FIG1} \end{center}
\end{figure}

\begin{figure}[ht]
\begin{center}
\includegraphics[angle=-90,width=.9\textwidth]{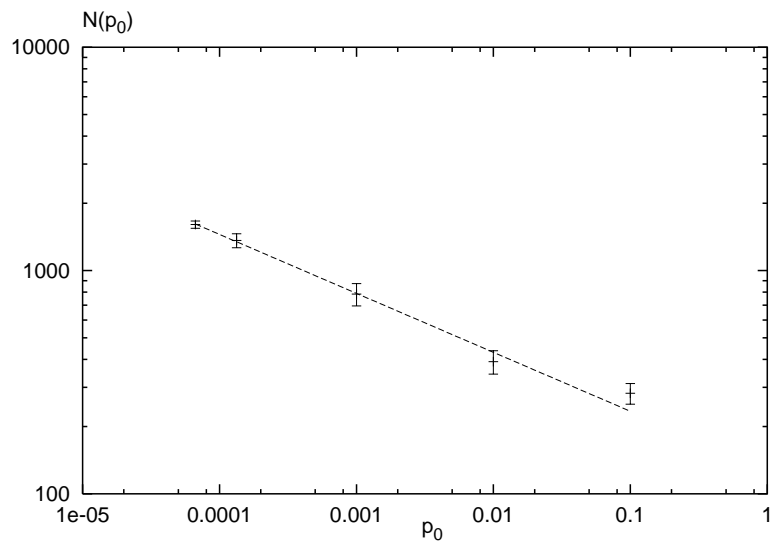}
\caption{The number of cycles $N$ needed to obtain the saturation $p=1$
as dependent on the initial density $p_0$.} \label{FIG2} \end{center}
\end{figure}

\begin{figure}[ht]
\begin{center}
\includegraphics[angle=-90,width=.9\textwidth]{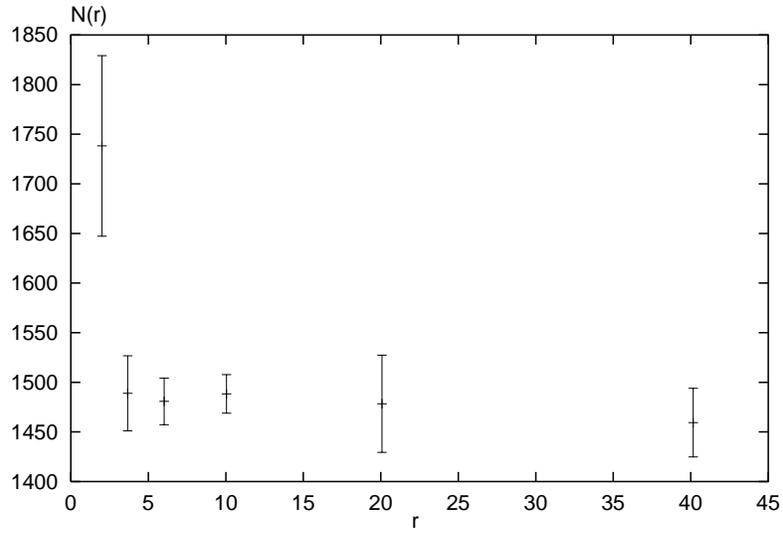}
\caption{The number of cycles $N$ needed to obtain the saturation $p=1$
as dependent on the ratio $r=L/d$.} \label{FIG3}
\end{center}
\end{figure}

\begin{figure}[ht]
\begin{center}
\includegraphics[angle=-90,width=.9\textwidth]{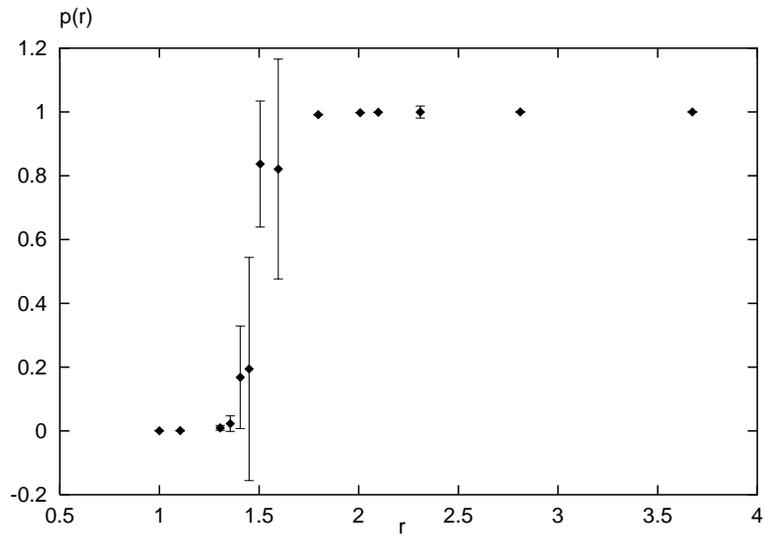}
\caption{The final occupation $p$ as dependent on the ratio $r=L/d$.}
\label{FIG4} \end{center}
\end{figure}

\begin{figure}[ht]
\begin{center}
\includegraphics[angle=-90,width=.9\textwidth]{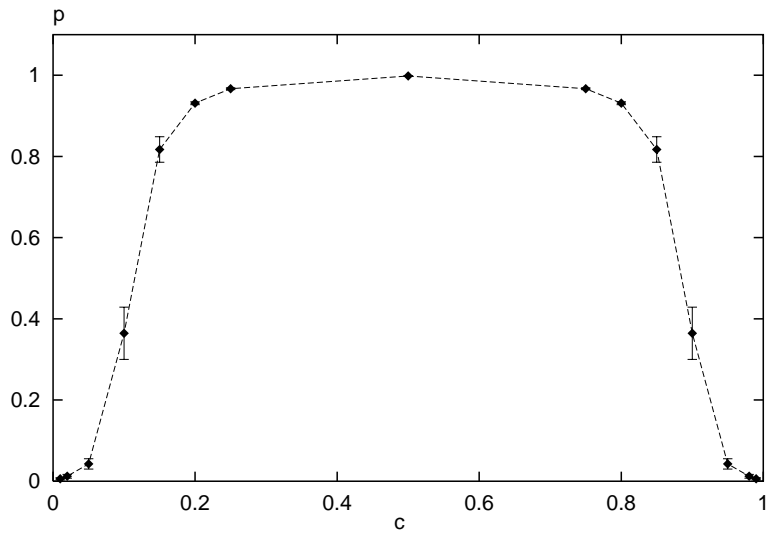}
\caption{The final occupation $p$ as dependent on the ratio $c$.}
\label{FIG5} \end{center}
\end{figure}

\end{document}